\documentclass[pra,aps,twocolumn, nobalancelastpage, superscriptaddress]{revtex4-2}
\usepackage[utf8]{inputenc}
\usepackage[utf8]{inputenc}
\usepackage{bbold}
\usepackage{geometry}
\usepackage{float}
\usepackage{graphicx}
\graphicspath{ {images/} }
\usepackage{braket}
\usepackage{subfigure}% in preamble
\usepackage[normalem]{ulem}
\usepackage{tabularx,colortbl,xcolor}
\usepackage[T1]{fontenc}
\usepackage{gensymb}
\usepackage{amsmath,amsfonts,amssymb}
\usepackage{amsmath}
\usepackage{amsthm}
\usepackage[]{natbib}
\usepackage{comment}
\newcommand{\be}{\begin{equation}}
\newcommand{\ee}{\end{equation}}
\newcommand{\ba}{\begin{aligned}}
\newcommand{\ea}{\end{aligned}}

\newcommand{\bc}{\begin{center}}
\newcommand{\ec}{\end{center}}
\newcommand{\beq}{\begin{equation}}
\newcommand{\eeq}{\end{equation}}
\newcommand{\beqq}{\begin{equation*}}
\newcommand{\eeqq}{\end{equation*}}
\newcommand{\beqa}{\begin{align}}
\newcommand{\eeqa}{\end{align}}
\newcommand{\barr}{\begin{array}}
\newcommand{\earr}{\end{array}}
\newcommand{\bi}{\begin{itemize}}
\newcommand{\ei}{\end{itemize}}

\usepackage[utf8]{inputenc}
\usepackage[T1]{fontenc} %fontenc makes the pdf ugly
\usepackage{lmodern} % Nicer font fith T1 fontenc
\usepackage{microtype} % Even nicer typography
\usepackage{graphicx}
\usepackage{dcolumn}
\usepackage{enumerate,amsthm,amssymb,color}
\usepackage{amsfonts}
\usepackage{amsmath}
\usepackage{mathtools}
\usepackage{dsfont}
\usepackage{siunitx}

\usepackage{subcaption}
\usepackage{ragged2e}
\DeclareCaptionJustification{justified}{\justifying}
\usepackage{bbm}
\usepackage{bm}
\usepackage{synttree}
\usepackage{float}
\usepackage{bbold}
\usepackage{braket}
\usepackage{xcolor}
\usepackage{gensymb}

\usepackage{booktabs}
\usepackage{upgreek, eurosym}
\usepackage[colorlinks=true]{hyperref}

\renewcommand\arraystretch{1.8}

\newcommand{\ketbra}[2]{|#1\rangle\!\langle#2|}

\graphicspath{{figures/}}

\begin{document}

%-------------------------------------------------------

\bibliographystyle{apsrev}

%-------------------------------------------------------

\title{Quantum Teleportation with Telecom Photons from Remote Quantum Emitters}

\author{Tim~Strobel}
\email{t.strobel@ihfg.uni-stuttgart.de}
\affiliation{Institut f\"ur Halbleiteroptik und Funktionelle Grenzfl\"achen, \\Center for Integrated Quantum Science and Technology (IQ\textsuperscript{ST}) and SCoPE, \\University of Stuttgart, Allmandring 3, 70569 Stuttgart, Germany}%

\author{Michal~Vyvlecka}
\affiliation{Institut f\"ur Halbleiteroptik und Funktionelle Grenzfl\"achen, \\Center for Integrated Quantum Science and Technology (IQ\textsuperscript{ST}) and SCoPE, \\University of Stuttgart, Allmandring 3, 70569 Stuttgart, Germany}%

\author{Ilenia~Neureuther}
\affiliation{Institut f\"ur Halbleiteroptik und Funktionelle Grenzfl\"achen, \\Center for Integrated Quantum Science and Technology (IQ\textsuperscript{ST}) and SCoPE, \\University of Stuttgart, Allmandring 3, 70569 Stuttgart, Germany}%

\author{Tobias~Bauer}
\affiliation{Fachrichtung Physik, Universit\"at des Saarlandes, Campus E2.6, 66123 Saarbr\"ucken, Germany}%

\author{Marlon~Sch\"afer}
\affiliation{Fachrichtung Physik, Universit\"at des Saarlandes, Campus E2.6, 66123 Saarbr\"ucken, Germany}%

\author{Stefan~Kazmaier}
\affiliation{Institut f\"ur Halbleiteroptik und Funktionelle Grenzfl\"achen, \\Center for Integrated Quantum Science and Technology (IQ\textsuperscript{ST}) and SCoPE, \\University of Stuttgart, Allmandring 3, 70569 Stuttgart, Germany}%

\author{Nand~Lal~Sharma}
\affiliation{Institute for Integrative Nanosciences, Leibniz IFW Dresden, Helmholtzstraße 20, 01069 Dresden, Germany}%

\author{Raphael~Joos}
\affiliation{Institut f\"ur Halbleiteroptik und Funktionelle Grenzfl\"achen, \\Center for Integrated Quantum Science and Technology (IQ\textsuperscript{ST}) and SCoPE, \\University of Stuttgart, Allmandring 3, 70569 Stuttgart, Germany}%

\author{Jonas~H.~Weber}
\affiliation{Institut f\"ur Halbleiteroptik und Funktionelle Grenzfl\"achen, \\Center for Integrated Quantum Science and Technology (IQ\textsuperscript{ST}) and SCoPE, \\University of Stuttgart, Allmandring 3, 70569 Stuttgart, Germany}%

\author{Cornelius~Nawrath}
\affiliation{Institut f\"ur Halbleiteroptik und Funktionelle Grenzfl\"achen, \\Center for Integrated Quantum Science and Technology (IQ\textsuperscript{ST}) and SCoPE, \\University of Stuttgart, Allmandring 3, 70569 Stuttgart, Germany}%

\author{Weijie~Nie}
\affiliation{Institute for Integrative Nanosciences, Leibniz IFW Dresden, Helmholtzstraße 20, 01069 Dresden, Germany}%

\author{Ghata~Bhayani}
\affiliation{Institute for Integrative Nanosciences, Leibniz IFW Dresden, Helmholtzstraße 20, 01069 Dresden, Germany}%

\author{Caspar~Hopfmann}
\thanks{Current affiliation: TU Dresden, Germany}
\affiliation{Institute for Integrative Nanosciences, Leibniz IFW Dresden, Helmholtzstraße 20, 01069 Dresden, Germany}%

\author{Christoph~Becher}
\affiliation{Fachrichtung Physik, Universit\"at des Saarlandes, Campus E2.6, 66123 Saarbr\"ucken, Germany}%

\author{Peter~Michler}
\affiliation{Institut f\"ur Halbleiteroptik und Funktionelle Grenzfl\"achen, \\Center for Integrated Quantum Science and Technology (IQ\textsuperscript{ST}) and SCoPE, \\University of Stuttgart, Allmandring 3, 70569 Stuttgart, Germany}%

\author{Simone~Luca~Portalupi}
\affiliation{Institut f\"ur Halbleiteroptik und Funktionelle Grenzfl\"achen, \\Center for Integrated Quantum Science and Technology (IQ\textsuperscript{ST}) and SCoPE, \\University of Stuttgart, Allmandring 3, 70569 Stuttgart, Germany}%

%

%-------------------------------------------------------

%

%-------------------------------------------------------

\begin{abstract}

The quest for a global quantum internet~\cite{Kimble2008,Lu2021} is based on the realization of a scalable network which requires quantum hardware with exceptional performance. Among them are quantum light sources providing deterministic, high brightness, high-fidelity entangled photons and quantum memories with coherence times in the millisecond range and above. To operate the network on a global scale, the quantum light source should emit at telecommunication wavelengths with minimum propagation losses. A cornerstone for the operation of such a quantum network is the demonstration of quantum teleportation~\cite{Cacciapuoti2020}. 
Here we realize full-photonic quantum teleportation employing one of the most promising platforms, i.e. semiconductor quantum dots, which can fulfill all the aforementioned requirements. Two remote quantum dots are used, one as a source of entangled photon pairs and the other as a single-photon source. The frequency mismatch between the triggered sources is erased using two polarization-preserving quantum frequency converters, enabling a Bell state measurement at telecommunication wavelengths. A post-selected teleportation fidelity of up to \num{0.721\pm0.033} is achieved, significantly above the classical limit, demonstrating successful quantum teleportation between light generated by distinct sources. These results mark a major advance for the semiconductor platform as a source of quantum light fulfilling a key requirement for a scalable quantum network. This becomes particularly relevant after the seminal breakthrough of addressing a nuclear spin in semiconductor quantum dots~\cite{Zaporski2023,Millington-Hotze2024} demonstrating long coherence times, thus fulfilling another crucial step towards a scalable quantum network.

\end{abstract}

%-------------------------------------------------------

\maketitle

\large

\noindent\textbf{Introduction}

\normalsize

In recent years, significant efforts have been made towards realizing the ambitious vision of a global quantum internet~\cite{Kimble2008,Lu2021} that would allow to securely connect distant nodes, and to interface with remote quantum computers\cite{Barz2012} or deployed quantum sensors. Fundamental for such a realization are quantum memories, to store and actively retrieve quantum information, and sources of quantum light to provide interconnection among different nodes. Several material systems are currently under investigation for their role in future quantum communication: atoms and ions~\cite{VanLeent2022,Langenfeld2021,Hartung2024,Kucera2024}, defect centers in diamond \cite{Pfaff2014,Hermans2022,Knaut2024,Stolk2024}, parametric processes \cite{Liu:PRL23,Ren:2017,Liao2017}, and semiconductor quantum dots (QDs)~\cite{Gao2013}. Because of recent milestone achievements demonstrating a tremendous improvement of spin coherence~\cite{Zaporski2023,Millington-Hotze2024}, epitaxial QDs showed great promise to function as quantum memory in future quantum networks. This is particularly appealing since they can effectively be interfaced with light generated by other QDs, which are known as efficient sources of single~\cite{Tomm2021,Ding2023,Wang2019,Somaschi2016,Uppu2020} and entangled-photons~\cite{DaSilva2021,Hopfmann2021,Huber2018,Pennacchietti2024}. Recent studies also showed all-photonic schemes with indistinguishable photons in cluster states as memory-free alternatives~\cite{Cogan2023,Coste2023}.

A key resource in quantum communication is quantum teleportation~\cite{Bouwmeester1997}, ideally realized with photons generated by remote sources of quantum light. Earlier studies with single QD emitters demonstrated their potential in teleportation experiments~\cite{Nilsson2013,Gao2013,Anderson2020,BassoBasset2021}. For successful implementation, photons capable of quantum interference and a high degree of entanglement are required. Furthermore, on-demand sources would be highly beneficial to upscale the network complexity, particularly when the generation process of single and entangled photons is deterministic and not probabilistic~\cite{hovsak2021effect,Bozzio2022}. Moreover, the ability to tune emitter wavelengths to a common target wavelength is essential for ensuring photon indistinguishability between distant sources. Two-photon interference with light generated by distinct sources has been investigated~\cite{Flagg2010,Patel2010,Weber2018,Gold2014,Giesz2015,Reindl2017,Weber2019} with recently reported high values of interference visibility~\cite{Zhai2022}. In addition, if long-distance propagation needs to be achieved, employing standard optical silica fibers for connecting distant nodes is highly beneficial. Indeed, silica fibers already represent the backbone of global telecommunication infrastructure, where light at telecommunication wavelengths experiences minimal propagation losses and limited photon wavepacket dispersion. Such behavior is even more crucial for quantum light. While low loss would allow for a reduction in the number of required repeater stations, low dispersion would ensure high interference visibility for photons over channels with different lengths~\cite{Weber2019}. These advantages render quantum light at telecommunication wavelengths particularly appealing for future implementation of quantum communication~\cite{vanLoock2020}. Despite the ongoing developments in realizing sources of quantum light operating at telecom wavelengths~\cite{Nawrath2023,Joos2024,Holewa2024,Barbiero2024,Lettner2021}, state-of-the-art performances are still set by the QDs emitting at near-infrared (NIR) wavelengths. For this reason, the use of frequency conversion was found to be an appealing approach to bridge this wavelength gap, and it was shown to be a powerful method for fine-tuning remote sources to the same wavelength, enabling two-photon interference~\cite{Weber2019, You2022}. Recently, quantum frequency converters designed to operate with QD light and capable of preserving the polarization state during conversion have been demonstrated~\cite{vanLeent2020,Strobel2024}.

Here we make use of epitaxially grown droplet etching semiconductor GaAs QDs to implement, for the first time, a full-photonic quantum teleportation experiment employing two distinct semiconductor QD sources of triggered quantum light. The single-photon state generated by one source is teleported onto the second, non-interfering, entangled photon emitted by the second QD, by performing a Bell state measurement (BSM). Photon interference is achieved by using two independent polarization-preserving quantum frequency converters to eliminate the photons' wavelength mismatch, fully preserving the high degree of entanglement during conversion~\cite{Strobel2024}. Appealing for out-of-the-lab experiments, the converted photons exhibit telecommunication wavelengths, therefore enabling long-distance propagation along standard silica fibers. Furthermore, we study the photon teleportation dynamics via time-resolved measurements in detail. An average teleportation fidelity of up to \num{0.721\pm0.033}, well above the classical limit, conclusively proves the success of teleportation in this full-photonic scheme. All the experimental results are confirmed by theoretical modeling, which explains each observed state's behavior in the teleportation process and enables a clear quantification of the envisioned possibilities in upcoming experiments.

\medskip

\large

\noindent\textbf{Remote Teleportation with Solid-State Quantum Emitters}

\normalsize
\begin{figure*}
    \centering
	\includegraphics[width=1\linewidth]{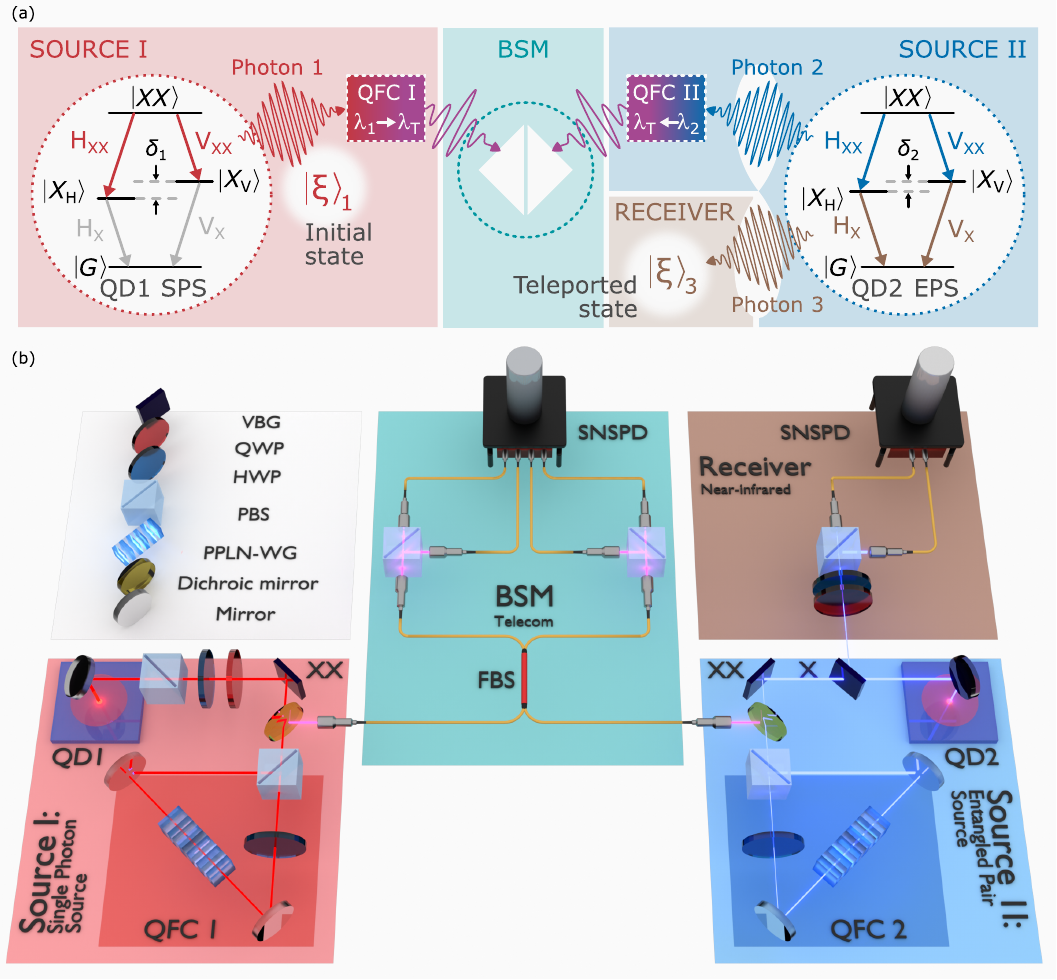}
	\caption{\textbf{Quantum teleportation setup:} (a) schematic of the experiment where QD$1$ is used as a single-photon source (SPS), while QD$2$ is used as an entangled pair source (EPS). Two independent quantum frequency converters are employed to convert the biexciton photons to a common telecommunication wavelength. After the Bell state measurement (BSM), the state of the single photon (named $\ket{\xi}_1$) is teleported onto the non-interfering exciton photon. (b) $3$D sketch of the setup: QD$1$ generates a single biexciton photon which is prepared in $\ket{\xi}_1$, frequency converted to telecommunication wavelength using polarization-preserving quantum frequency conversion (QFC), and sent to the fiber beamsplitter (FBS) for the BSM. QD$2$ generates an entangled photon pair: while the exciton photon is sent to the near-infrared receiver, the biexciton is frequency converted to match the wavelength of the converted biexciton photon of QD$1$. Polarizing beamsplitters (PBS) are used in the BSM and receiver side.}
    \label{fig:fig1} 
\end{figure*}
\noindent
\noindent Figure~\ref{fig:fig1}a shows a general schematic of the experiment. Two remote QDs are used: QD$1$ functions as a single-photon source (SPS), emitting Photon~1. QD$2$ serves as an entangled pair source (EPS), emitting an entangled photon pair: Photon~2 and Photon~3. In both cases, the QDs are excited via pulsed two-photon excitation~\cite{Akopian2006, Young2006, Hafenbrak2007} and generate photons via the biexciton-exciton cascade ($\ket{XX}\rightarrow\ket{X}\rightarrow\ket{G}$)~\cite{MichlerPortalupi2024}. The biexciton (XX) photons (Photon~1 and Photon~2) are sent to a BSM setup, after two distinct quantum frequency conversion (QFC) processes (see methods and \cite{Strobel2024}), while the exciton (X) emission of QD2 (Photon~3) is analyzed. The joint BSM projects Photon~1 and Photon~2 onto a maximally entangled Bell state teleporting the polarization state of Photon~1 (named $\ket{\xi}_1$) onto Photon~3. The receiver reconstructs the polarization state of the Photon~3 (named $\ket{\xi}_3$) conditioned on the BSM result.\\

\begin{figure}
    \centering
	\includegraphics[width=1\linewidth]{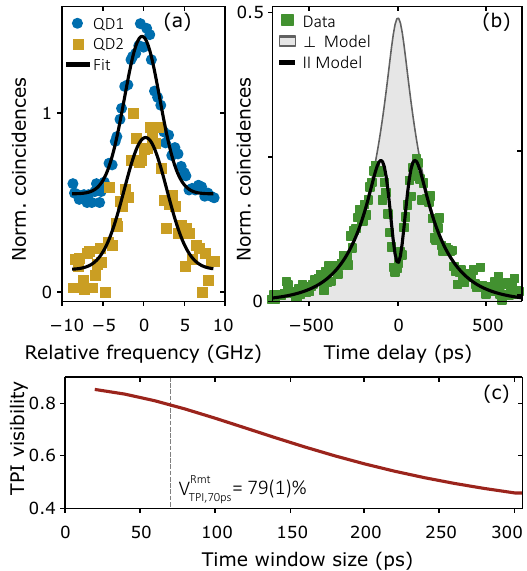}%
	\caption{\textbf{Linewidth and interference of remote quantum light sources}: (a) High-resolution linewidth measurements of the QDs biexciton emissions ($\ket{H}$ component) after quantum frequency conversion, recorded with a Fabry-P\'{e}rot interferometer, also depicting their spectral overlap. The data points depict the measured results and the solid lines a Gaussian fit function. The data are shifted in the y-direction for clarity. (b) The center peak of a remote two-photon-interference (TPI) experiment between Photon~1 and Photon~2 with parallel polarization (green squares) is shown (see supplement for full histogram). A model function as proposed by \cite{legero2003time} is used to fit these data (solid black line). The orthogonal polarization setting (gray area) is deduced from the fit results of the parallel setting. (c) Temporally post-selected remote two-photon-interference visibility calculated from the fit results given in (b). The time window is centered around zero time delay.}
    \label{fig:fig2}
\end{figure}

Figure~\ref{fig:fig1}b depicts a detailed illustration of the experimental configuration. During the experimental procedure, a pulsed laser coherently prepares a XX state in two epitaxially grown droplet etching GaAs QDs~\cite{DaSilva2021,Hopfmann2021} remotely located in different cryostats. The excitation is followed by a cascaded emission of two polarization-entangled photons at NIR wavelengths ($\sim\SI{780}{\nano\meter}$). The polarization state $\ket{\xi}_1$ of Photon~1 (to be teleported) is prepared by sending it through a polarizing beamsplitter (PBS) followed by half- (HWP) and quarter-wave plate (QWP). Photon~2 and Photon~3 share a maximally entangled oscillating state $1/\sqrt{2}\left(\ket{HH}_{2,3}+e^{i\delta_2t/\hbar}\ket{VV}_{2,3}\right)$, where $\ket{H}$ ($\ket{V}$) represents horizontal (vertical) polarization, $\delta_2$ the fine-structure splitting (FSS) of the EPS, $t$ the time between XX and X emission, and $\hbar$ the reduced Planck constant. For slow FSS-induced oscillations (here $\delta_2=\SI{2.1}{\micro\electronvolt}$) relative to the emitter decay time the latter state can be simplified to a maximally entangled Bell state $\ket{\Phi^+}_{2,3}=1/\sqrt{2}\left(\ket{HH}_{2,3}+\ket{VV}_{2,3}\right)$ (for more details see supplement). The wavelengths of Photon~1 and Photon~2 at \SI{\sim780}{\nano\meter} do not spectrally overlap, prohibiting interference. To enable two-photon interference required for a successful BSM, polarization-preserving QFC is employed (see Fig.~\ref{fig:fig1}b and methods)~\cite{vanLeent2020}. This process converts the XX photons (Photon~1 and Photon~2) to a common telecommunication wavelength (\SI{1515}{\nano\meter}, see Fig.~\ref{fig:fig2}a) leaving their quantum state unaltered as shown in~\cite{Strobel2024}. Being at technologically relevant telecommunication wavelengths also opens the way for prospective long-distance teleportation experiments. After interference, the (three-photon) state can be written in the Bell basis:

\begin{equation}
    \begin{split}
        \ket{\Psi_{\text{tot}}} &= \ket{\xi}_1 \otimes \ket{\Phi^+}_{2,3} = \frac{1}{2} \left( \ket{\Phi^+}_{1,2}\ket{\xi}_3 
        + \ket{\Phi^-}_{1,2}\sigma_3\ket{\xi}_3 \right. \\
        & \quad \left. + \ket{\Psi^+}_{1,2}\sigma_1\ket{\xi}_3 
        - \ket{\Psi^-}_{1,2}\sigma_1\sigma_3\ket{\xi}_3 \right),
    \end{split}
\end{equation}
with Pauli matrices $\sigma_1$ and $\sigma_3$ (for exemplary calculations see supplement) and Bell states $\ket{\Phi^{\pm}}_{1,2}=1/\sqrt{2}\left(\ket{HH}_{1,2}\pm\ket{VV}_{1,2}\right)$ and $\ket{\Psi^{\pm}}_{1,2}=1/\sqrt{2}\left(\ket{HV}_{1,2}\pm\ket{VH}_{1,2}\right)$. Here, the BSM unit is implemented as a single-mode fiber beamsplitter (FBS) with a PBS at each output arm, followed by superconducting nanowire single-photon detectors (SNSPD). In this polarization selection BSM configuration the projections of Photon~1 and Photon~2 onto Bell states $\ket{\Psi^+}_{1,2}$ or $\ket{\Psi^-}_{1,2}$ can be identified, increasing the BSM efficiency considerably~\cite{VanLeent2022,BassoBasset2021}. A successful BSM heralds the unitary transformation to be applied to Photon~3 to reconstruct the initially prepared quantum state of Photon~1 (see supplement). To probe the teleported polarization state, Photon~3 is analyzed with a tomography unit consisting of a QWP, HWP followed by a PBS, and two SNSPDs. In the following the results of one projection on $\ket{\Psi^-}_{1,2}$ are discussed (other combinations can be found in the supplement).\\
\noindent

% New text that avoids jumping between indistinguishability and entanglement
Two key requirements for successful quantum teleportation are a high degree of entanglement of the EPS~\cite{BassoBasset2021} and a high indistinguishability between the two XX photons entering the BSM.
The former is intrinsic to the employed QD structure and it is maintained by the QFC setup (entanglement fidelities up to \num{0.97} as shown in~\cite{Strobel2024}, see supplement). The latter is mainly achieved thanks to the erasure of the initial frequency mismatch between the remote interfering XX photons via the precise spectral tuning of the pump fields in the QFC processes~\cite{Weber2019} (being the interference visibility only eventually limited by the pumping scheme and spectral broadening). 

Figure~\ref{fig:fig2}a depicts a high-resolution spectrum of the XX emission lines of QD$1$ and QD$2$ at telecommunication wavelength after frequency matching during QFC. A Gaussian fit function provides a linewidth of \SI{4.3\pm0.1}{\giga\hertz} (\SI{5.2\pm0.4}{\giga\hertz}) for QD$1$ (QD$2$). The two given lines have a relative spectral offset of \SI{0.43\pm0.27}{\giga\hertz}, due to pump laser drifts in the QFC setup. Decay-time measurements ($\tau_{\text{XX}}^{\text{QD}1,2}=\SI{120}{\pico\second},\,\SI{176}{\pico\second}$) suggest the photon Fourier limit to be at \SI{1.322\pm0.005}{\giga\hertz} and \SI{0.903\pm0.010}{\giga\hertz}. The deviation of the measured linewidth from the Fourier limit is caused by inhomogeneous spectral broadening mechanisms~\cite{Vural2020,MichlerPortalupi2024} and suffices for a Gaussian approximation in the fit. The indistinguishability of the two converted emissions is probed by a two-photon-interference (TPI) experiment at the outputs of the FBS with linear polarized photons. The central peak of the correlation measurement is shown as green data points in Fig.~\ref{fig:fig2}b. From this, the remote two-photon-interference visibility $V_{\text{TPI}}^{\text{Rmt}}$ is evaluated. While this central peak is expected to vanish for fully indistinguishable photons, in the present case the interference visibility is limited to \SI{30\pm1}{\percent}. This has two reasons: first, the time-ordered cascade of the three-level system gives an upper limit set by the XX and X decay rates $V_{\text{TPI,max}}^{\text{Rmt}}=\gamma_{XX}/(\gamma_{XX}+\gamma_X)=\SI{59}{\percent}$~\cite{scholl2020crux}, and second, the inhomogeneous spectral broadening mechanisms discussed before. The interference visibility can be increased through temporal post-selection, which can mitigate the impact of the two aforementioned mechanisms. Figure~\ref{fig:fig2}c depicts the interference visibility for an increasing time window centered around zero time delay. Indeed, the visibility of $V_{\text{TPI,\SI{70}{\pico\second}}}^{\text{Rmt}}=\SI{79\pm1}{\percent}$ found for a time window of \SI{70}{\pico\second} (minimal post-selection time window in the teleportation experiment discussed below) drops to \SI{30\pm1}{\percent} without temporal post-selection.
\begin{figure*}
    \centering
	\includegraphics[width=1\linewidth]{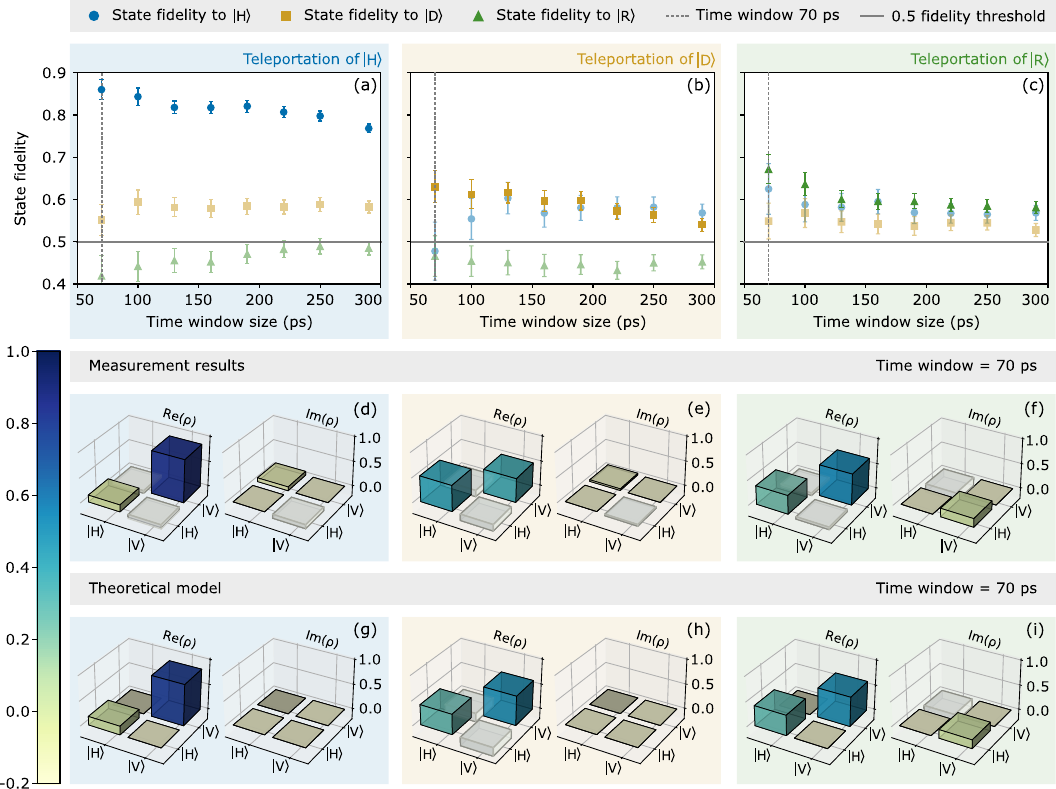}%

 	\caption{\textbf{Teleportation of three conjugate polarization states:} The teleportation experiment is repeated with three conjugate input states of Photon~1: $\ket{\xi}_1=\ket{H}$, $\ket{D}$ and $\ket{R}$ depicted in (a), (b) and (c). For all three experiments, the fidelity of the measured teleported state $\ket{\xi}_3$ (heralded by Bell state $\ket{\Psi^-}_{1,2}$) to the three polarization states $\ket{H}$, $\ket{D}$ and $\ket{R}$ is shown in blue, yellow and green. The error bars correspond to one standard deviation. The solid gray line at \num{1}/\num{2} symbolizes zero degree of polarization in the respective basis. (d), (e) and (f) ((g), (h) and (i)) present real and imaginary parts of the density matrix of the teleported state $\ket{\xi}_3$ at a \SI{70}{\pico\second} time window obtained from the measurement data (theoretical model) when teleporting $\ket{\xi}_1=\ket{H}$, $\ket{D}$ and $\ket{R}$ respectively (before the application of any unitary transformation). To obtain the theoretical density matrices (g), (h) and (i), the following parameters were assumed in the theory model: mode overlap $M=0.85$, dephasing time $T_2=\SI{35}{\pico\second}$, cross-dephasing time $\tau_{HV}$=\SI{5}{\nano\second}, spin scattering time $\tau_{ss}=\SI{5}{\nano\second}$, decay time of exciton $\tau_X=\SI{171}{\pico\second}$, TPI visibility $V=\SI{79}{\percent}$ and ratio of true three-fold coincidences $k=0.85$ (see methods).}
    \label{fig:fig3}
\end{figure*}
\newline
\newline
To perform the all-photonic teleportation experiment, Photon~1 is prepared in three conjugate polarization states $\ket{\xi}_1=\ket{H}$, $\ket{D}$, and $\ket{R}$, respectively. We measure three-photon coincidences between the BSM and Photon~3 for time windows ranging from \SI{70}{\pico\second} to \SI{290}{\pico\second} (the latter being a trade-off between covering the entire interference peak and minimizing unwanted background coincidences), resulting in averaged coincidence rates between \SI{0.11\pm0.03}{\milli\hertz} and \SI{2.5\pm0.7}{\milli\hertz}, respectively. The density matrix of the teleported state is reconstructed from the coincidence measurements (see methods and supplement).\\
In Fig.\,\ref{fig:fig3}a-c the fidelity of the teleported state $\ket{\xi}_3$ (heralded by $\ket{\Psi^-}_{1,2}$) to the three conjugate input states is calculated. Because the TPI visibility decreases as a function of integration time as shown in Fig.~\ref{fig:fig2}c, the fidelity is expected to also depend on this time window. Therefore, the fidelity is evaluated for various three-photon coincidence time windows between \SI{70}{\pico\second} and \SI{290}{\pico\second}. The data points are the measurement results with error bars given by one standard deviation of a distribution obtained via a Monte-Carlo simulation (\num{10000} runs) assuming Poissonian statistics (see supplement). In an ideal scenario when teleporting $\ket{H}$ ($\ket{D}$ or $\ket{R}$) one would expect the fidelity to the $\ket{H}$ ($\ket{D}$ or $\ket{R}$) state $f^{\ket{H}\xrightarrow{}\ket{H}}$ ($f^{\ket{D}\xrightarrow{}\ket{D}}$ or $f^{\ket{R}\xrightarrow{}\ket{R}}$) to be unity and the respective remaining two fidelities $1/2$ (gray line in Fig.\,\ref{fig:fig3}a-c). For example, a fidelity to $\ket{H}$ of $1$ means the photon is maximally polarized in $\ket{H}$ and a fidelity to $\ket{R}$ of $1/2$ means the photon has no polarization component in the $\ket{R}$-$\ket{L}$-basis. In Fig.\,\ref{fig:fig3}a the fidelity of the teleported state $\ket{H}$ to state $\ket{H}$ is $f_{\SI{70}{\pico\second}}^{\ket{H}\xrightarrow{}\ket{H}}=\num{0.860\pm0.023}$ for a \SI{70}{\pico\second} time window. For longer time windows $f^{\ket{H}\xrightarrow{}\ket{H}}$ drops only slightly but stays above $0.7$. In Fig.\,\ref{fig:fig3}b (Fig.\,\ref{fig:fig3}c) $f_{\SI{70}{\pico\second}}^{\ket{D}\xrightarrow{}\ket{D}}=\num{0.630\pm0.038}$ ($f_{\SI{70}{\pico\second}}^{\ket{R}\xrightarrow{}\ket{R}}=\num{0.672\pm0.034}$) and drops to \num{0.55} (\num{0.6}) for longer time windows. All remaining fidelities of states conjugate to the initial state of Photon~1 show deviations of \num{\pm0.1} from $1/2$. The three described teleportation experiments are modeled with theoretical simulations (see methods and supplement for further details). Based on this model the two main contributors to the non-unity of fidelities $f^{\ket{H}\xrightarrow{}\ket{H}}$, $f^{\ket{D}\xrightarrow{}\ket{D}}$ and $f^{\ket{R}\xrightarrow{}\ket{R}}$ are the limited TPI visibility and multi-photon contributions from the QFC process (anti-Stokes Raman scattered photons generated at the target wavelength~\cite{Kuo2018}, see supplement). For larger time windows the TPI visibility drops (Fig.\,\ref{fig:fig2}c) while the fraction of background counts increases leading to the observed decrease in the fidelities. Higher teleportation fidelities of $f^{\ket{H}\xrightarrow{}\ket{H}}$ are a result of the BSM basis choice ($\ket{H},\ket{V}$) leading to additional classical correlations between the BSM and Photon~3 (see supplement for a detailed explanation). Examining the corresponding density matrix $\rho$ of the teleported state is essential to understanding the variations in conjugate fidelities. Figure\,\ref{fig:fig3}d, e, and f (g, h and i) depict the measured (simulated) density matrices of the three teleported states respectively for an exemplary time window of \SI{70}{\pico\second} around zero time delay, before application of any unitary transformation. The imbalance between the diagonal elements $\rho_{HH}$ and $\rho_{VV}$ in the real part of Fig.\,\ref{fig:fig3}f, h and i arise from classical correlations, favoring the teleportation of the $\ket{H}$ state due to differences in TPI visibilities for $\ket{H}$ and $\ket{V}$ wave packets (manifesting itself in a lowered $\rho_{HH}$ and increased $\rho_{VV}$ when projecting onto $\ket{\Psi^-}_{1,2}$). This imbalance in TPI visibilities comes from the QD's non-zero FSS, non-perfect temporal wavepacket overlap, and experimental setup birefringence, which can be described by the degree of polarization mode overlap $M_p$ (see methods). The measured off-diagonal elements of the real part $\rho_{HV}$ and $\rho_{VH}$ for the teleportation of $\ket{H}$ and $\ket{R}$ in Fig.\,\ref{fig:fig3}d,f are slightly lowered compared to the theoretical model in Fig.\,\ref{fig:fig3}g,i. For the teleportation of $\ket{D}$ non-zero off-diagonals are expected (Fig.\,\ref{fig:fig3}e,h). The off-diagonal elements are affected by decoherence processes within the QD, which are determined by factors such as non-zero FSS, cross-dephasing time $\tau_{HV}$, spin scattering time $\tau_{ss}$, and dephasing time $T_2$ (see supplement). Additionally, the small non-zero off-diagonal elements $\rho_{HV}$ and $\rho_{VH}$ in the imaginary part for the teleported states $\ket{H}$ and $\ket{D}$ (Fig.\,\ref{fig:fig3}d,e in comparison to Fig.\,\ref{fig:fig3}g,h) indicate an imperfect transformation between the QD bases and the measurement bases. These effects also result in the observed deviations from $1/2$ for fidelities to states conjugate to $\ket{\xi}_1$ in Fig.\,\ref{fig:fig3}a-c. Calculated fidelities between the measured and modeled density matrices above \SI{97}{\percent} confirm the agreement between the experiment and the theoretical model (see supplement). 
\begin{figure}
    \centering
	\includegraphics[width=1\linewidth]{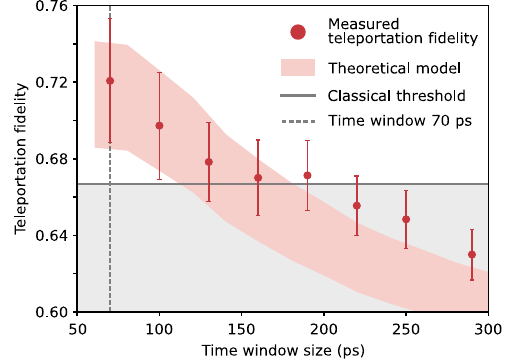}%
	\caption{\textbf{Average teleportation fidelity}: The average fidelity over all three performed teleportation experiments $\bar{f}=(f^{\ket{H}\xrightarrow{}\ket{H}}+f^{\ket{D}\xrightarrow{}\ket{D}}+f^{\ket{R}\xrightarrow{}\ket{R}})/3$, heralded by Bell state $\ket{\Psi^-}_{1,2}$ is shown. Red solid dots represent the measured results with error bars corresponding to one standard deviation. The red shaded area shows the theoretically modeled data and the gray solid line symbolizes the classical threshold of \num{2}/\num{3}. The uncertainty in the theoretical model arises from some parameters that cannot be precisely determined in our experimental setting. Therefore they were based on literature~\cite{Hudson:PRL07} estimated within the following intervals to account for their variability: polarization mode overlap $M_p=\left[ 0.8, 0.9\right]$, cross-dephasing time $\tau_{HV}=\left[ 1, 10\right]\si{\nano\second}$, spin scattering time $\tau_{ss}=\left[ 1, 10\right]\si{\nano\second}$. The other parameters are assumed as: dephasing time $T_2=\SI{35}{\pico\second}$ and decay time of exciton $\tau_X=\SI{171}{\pico\second}$.}
    \label{fig:fig4}
\end{figure} 
\noindent
From the results in Fig.\,\ref{fig:fig3}a-c an average teleportation fidelity, denoted as $\bar{f}=(f^{\ket{H}\xrightarrow{}\ket{H}}+f^{\ket{D}\xrightarrow{}\ket{D}}+f^{\ket{R}\xrightarrow{}\ket{R}})/3$ is determined. If one would repeat the described teleportation experiment with every possible state on the Poincar\'e sphere, the expected average teleportation fidelity is given by $\bar{f}$. For this reason, the average teleportation fidelity is the figure of merit for this work. The value of $\bar{f}$ indicates successful quantum teleportation when it exceeds the classical threshold of $2/3$. Solid dots (the red shaded area) in Fig.\,\ref{fig:fig4} show the average teleportation fidelity of the experimental (modeled) teleportation process, with error bars obtained via an error propagation of standard deviations. The data corresponds to the state heralded by $\ket{\Psi^-}_{1,2}$. For a time window of \SI{70}{\pico\second} the measured average teleportation fidelity $\bar{f}_{\SI{70}{\pico\second}}$ equals \num{0.721\pm0.033}, being \num{1.6} standard deviations above the classical threshold. The fidelity stays above this threshold up to \SI{190}{\pico\second}, which is longer than the employed XX photon decay time. For longer time windows this value drops below the classical threshold reaching a steady state at \num{0.63\pm0.012}. Between \SI{70}{\pico\second} and \SI{160}{\pico\second} the modeled results closely reproduce the experimental findings. A divergence between the experiment and model, still within one standard deviation, occurs for longer time windows. Other BSM combinations can be found in the supplement.\\

\begin{figure}
    \centering
	\includegraphics[width=1\linewidth]{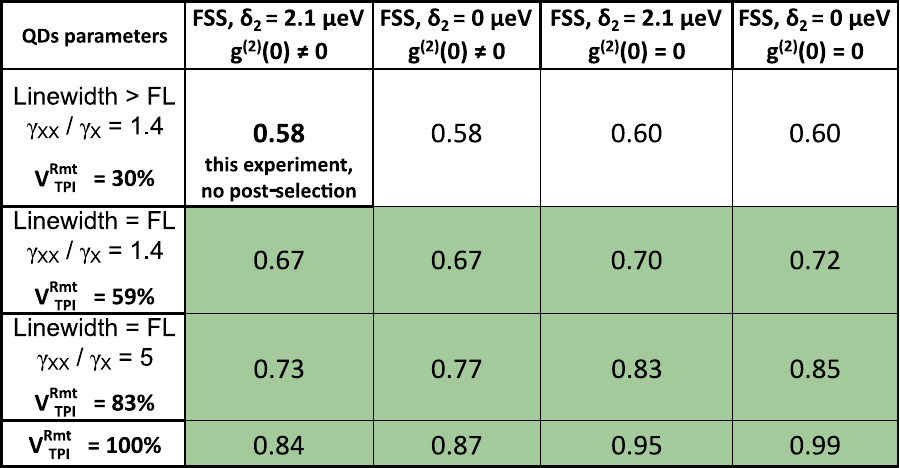}%

 	\caption{\textbf{Theoretical average teleportation fidelity as a function of linewidth, XX and X decay rates, FSS, and $\bm{g^{(2)}(0)}$ without temporal post-selection:} first slot, current experimental parameters, with a TPI visibility of \SI{30}{\percent}, consistent with experimental results. Each row shows the impact of FSS and $g^{(2)}(0)$. TPI visibility can increase to $V_{\text{TPI,max}}^{\text{Rmt}}=\SI{59}{\percent}$ if Fourier-limited linewidths are considered. The radiative cascade still limits this value, nevertheless, average teleportation fidelities above the classical limit can be expected. Accelerating the biexciton decay selectively (so that $\gamma_{XX}$/$\gamma_{X} = 5$) can increase the TPI visibility to \SI{83}{\percent}. The last row shows the achievable fidelities for $V_{\text{TPI,max}}^{\text{Rmt}}=\SI{100}{\percent}$ as a function of FSS and $g^{(2)}(0)$. The following parameters were used: decay time of exciton $\tau_X=\SI{171}{\pico\second}$, polarization mode overlap $M_p=1$; when present, dephasing time $T_2=\SI{35}{\pico\second}$, cross-dephasing time $\tau_{HV}=\SI{10}{\nano\second}$, spin scattering time $\tau_{ss}=\SI{10}{\nano\second}$, ratio of wanted three-fold coincidences to all detected three-fold coincidence $k=0.85$.}
    \label{fig:fig5}
\end{figure}

The results achieved show successful teleportation when temporal post-selection is applied. The requirement of post-selection is mainly due to the intrinsic performance of the utilized sources and excitation method. Thanks to the developed theory, it is possible to quantify the impact of the experimental parameters on the average teleportation fidelity such as TPI visibility, FSS, and single-photon purity. The average teleportation fidelity achievable with optimal QD parameters is calculated to highlight the significance of these factors, see Fig.~\ref{fig:fig5}. Among them, TPI visibility was identified as the most crucial parameter, which must be optimized for further improvement of the average teleportation fidelity. In the case of Fourier-limited sources (as shown e.g. for gated structures~\cite{Zhai2022}), where the visibility of the used QD is still limited by the decay rates of the radiative cascade to $V_{\text{TPI,max}}^{\text{Rmt}}=\SI{59}{\percent}$ (as discussed before), the calculated fidelity surpasses the classical limit across all scenarios without requiring temporal filtering. Further improvements in TPI visibility can be achieved by incorporating a photonic structure that selectively accelerates the XX decay~\cite{bauch2024demand}. Assuming $\gamma_{XX}=5\gamma_{X}$, which corresponds to a visibility of $\SI{83}{\percent}$, leads to an average teleportation fidelity of \num{0.73}. Reducing the FSS and $g^{(2)}(0)$ further improves the achievable fidelity. The former can be minimized by mechanical strain~\cite{Huber2018} or exploiting the Stark effects~\cite{Chen2024}. The latter can be improved by reducing the conversion-related noise with narrower spectral filtering. In such a scenario ($\text{FSS}=0$, $g^{(2)}(0)=0$), an average teleportation fidelity of \num{0.85} could be attained. Assuming unity interference visibility, a fidelity above \num{0.8} is achieved in every scenario, reaching up to \num{0.99}.

\medskip

\large

\noindent\textbf{Discussion}
\normalsize
\noindent
\noindent
\noindent Here, we presented the experimental demonstration of an all-photonic quantum teleportation experiment using distinct semiconductor quantum emitters. The frequency mismatch between the remotely emitted photons is erased by employing quantum frequency conversion. Furthermore, this allows for converting the interfering photons to telecommunication wavelengths, a necessary step for upcoming long-distance implementations. These results demonstrate the maturity of quantum dot-based technology, showing an important building block for future quantum communication, i.e. the successful teleportation of a photon state on one photon of a polarization-entangled pair. Employing a polarization-selective BSM setup, an average teleportation fidelity of up to $\bar{f}_{\SI{70}{\pico\second}}=\num{0.721\pm0.033}$, well above the classical limit, is measured for a temporal post-selection window of \SI{70}{\pico\second}. The paramount factor in this realization is the capability of the QFC process to fully preserve the polarization state of the photons involved in the conversion processes. All results are fully explained by the introduced theoretical model. We foresee that a drastic improvement in teleportation fidelity is within reach when employing tailored, optimized sample designs. This will be key for the future implementation of more complex experiments such as entanglement swapping, and the realization of more advanced quantum communication architectures.

\medskip

\large

\noindent\textbf{Methods}

\normalsize

\noindent\textbf{Two-Photon Excitation and employed Quantum Light Sources} The samples used in this work consist of droplet etching GaAs QDs~\cite{DaSilva2021}. Their high in-plane symmetry is reflected in low FSS. A dielectric antenna device is realized to enhance the light extraction efficiency: the semiconductor membrane embedding the QDs is glued on the bottom of a millimeter-scale GaP lens~\cite{Hopfmann2021}. Below the membrane, a gold layer, acting as a bottom mirror, is deposited. The careful control of all layers' thicknesses allows for improved extraction and a narrow far-field emission profile (see supplement). Additionally, the lens enables a tighter focusing of the excitation laser, reducing the power necessary to reach population inversion ($\pi$-pulse). Optical excitation of the biexciton state is performed via pulsed resonant two-photon excitation (TPE)~\cite{Strobel2024}: this allows for the preparation of the biexciton with \SI{45\pm7}{\percent} fidelity, a value estimated via the observed Rabi oscillations (see supplement). Pulsed excitation (with a repetition rate of \SI{304.8}{\mega\hertz}) results in triggered emission of entangled photon pairs via the biexciton-exciton cascade~\cite{MichlerPortalupi2024}. For QD$1$, an FSS of $\delta_1=\SI{10}{\micro\electronvolt}$ is found and the biexciton photon is directly projected onto the state $\ket{\xi}_1$ that will be successively teleported. The second quantum dot, QD$2$, is also excited under TPE, has an FSS of $\delta_2=\SI{2.1}{\micro\electronvolt}$, and the emitted photon pair has a concurrence of \num{0.9544\pm0.0002}~\cite{Strobel2024}. More information on count rates can be found in the supplement. It is worth mentioning that the polarization-preserving quantum frequency converters (detailed below) allow for maintaining a high degree of entanglement for QD$2$, as well as a well-controlled polarization on the state $\ket{\xi}_1$ of QD$1$ which is prepared to $\ket{H}$, $\ket{D}$ and $\ket{R}$ before conversion. Two separate closed-cycle cryostats (Cryostation s50 with Cryo-Optic configuration, from Montana Instruments) are employed to keep the samples at \SI{6}{\kelvin} for the full duration of the experiments.

\noindent\textbf{Quantum Frequency Conversion} The quantum frequency converters employed in this experiment, previously used in~\cite{vanLeent2020, VanLeent2022, Strobel2024}, utilize difference frequency generation in periodically poled lithium niobate waveguides. In this process, QD photons are mixed with a strong pump laser to produce telecommunication wavelength photons, using a Sagnac-type setup to facilitate polarization-preserving conversion. To enable the Bell state measurement, the photons from both QDs at \SI{779.90\pm0.01}{\nano\meter} (respectively, \SI{780.00\pm0.01}{\nano\meter}) are converted to a common telecommunication wavelength of \SI{1515.53\pm0.04}{\nano\meter} by using separate pump lasers at \SI{1606.74}{\nano\meter} (\SI{1607.1612}{\nano\meter}) for each conversion device. Raman-scattering of the pump field creates broadband noise around the target wavelength which is spectrally filtered using a bandpass filter (\SI{30}{\nano\meter} FWHM), and a \SI{25}{\giga\hertz} VBG, resulting in a \SI{50}{\kilo\hertz} noise count rate for each converter. Total device efficiencies are \SI{49}{\percent} and \SI{47}{\percent}. 

\noindent\textbf{Detection System} For the described experiments, a set of six superconducting nanowire single-photon detectors is utilized ($4$ designed for telecom C-band operation, $2$ for NIR, particularly at \SI{780}{\nano\meter}). The detectors are installed in two distinct cryostats: one operates the four telecom detectors, which have \SI{85}{\percent} detection efficiency, \SI{37}{\pico\second} time resolution (full width at half maximum), and \SI{300}{\hertz} dark count rate. The second cryostat operates the $780\,$nm detectors, which have \SI{85}{\percent} detection efficiency, \SI{44}{\pico\second} time resolution, and \SI{150}{\hertz} dark count rate. Both detector systems are the Eos model of Single Quantum. The simultaneous operation is enabled by Swabian Instruments time tagging unit (Time Tagger Ultra). The overall system time resolution per channel is \SI{41}{\ps} and \SI{48}{\ps} at telecommunication and NIR wavelength respectively.

\noindent\textbf{Polarization Control}
The alignment of the QD polarization basis (QD$2$) with the measurement polarization basis is carried out by maximizing the entanglement while varying the wave plate angles of the projection units. Birefringence caused by optical fibers is compensated (between measurement runs) by sending horizontally, diagonally, or circularly polarized laser light through the setup. Monitoring the polarization at the output of the FBS in the BSM setup with a polarimeter allows for the correction of the birefringence using fiber polarization controllers and wave plates.

\noindent\textbf{Theory Model} 
The theoretical model relies on the quantum process matrix formalism for quantum teleportation with realistic QDs~\cite{BassoBasset2021,Hudson:PRL07}. This formalism provides an analytical description of the output state of the teleportation protocol, denoted as $\hat{\rho}_{\text{teleported}}(\hat{\rho}{_\text{in}})$, which depends on the input state $\hat{\rho}_{\text{in}}$ and incorporates the effects of limited single-photon purity, non-zero FSS, and other decoherence processes in the QD. Here, we corrected the original output state matrix $\hat{\rho}_{\text{teleported}}$ using a classical interference term that prioritizes the $\ket{H}$ state as the output from the teleportation process. The weight of this correction term is determined by the polarization mode overlap $M_p$, which accounts for the spectral and spatial distinguishability between the $\ket{H}$ and $\ket{V}$ wave packets generated by the QD due to non-zero FSS, imperfect polarization mode overlap at the FBS in the BSM caused by different temporal profiles of the interfering single photons and setup birefringence. In comparison to the standard mode overlap, which is characterized by TPI visibility $V$, the polarization mode overlap describes the difference between interference visibilities $\ket{H}$ and $\ket{V}$ polarization states. The output state is then given as the following state mixture

\begin{equation}
\hat{\rho}_{\text{out}}(\hat{\rho}_{\text{in}})=M_p\hat{\rho}_{\text{teleported}}(\hat{\rho}_{\text{in}})+(1-M_p)\ketbra{H}{H} \ ,
\end{equation}
where optimized TPI visibility for $\ket{H}$ polarization state was assumed (see supplement). The parameters of the used QDs were either determined by spectral and radiative decay measurements (decay time of exciton $\tau_X=\SI{171}{\pico\second}$, dephasing time $T_2=\SI{35}{\pico\second}$) or fixed based on the literature on similar QD systems (cross-dephasing time $\tau_{HV}=\left[ 1, 10\right]\si{\nano\second}$, spin scattering time $\tau_{ss}=\left[ 1, 10\right]\si{\nano\second}$)~\cite{Hudson:PRL07}. In such cases, the uncertainty interval was assumed as an input for the theoretical model, as the dephasing times $\tau_{HV}$ and $\tau_{ss}$ weren't measured for the QD used. The FSS reduces the polarization mode overlap, leading to spectral distinguishability of the $\ket{H}$ and $\ket{V}$ polarization wave packets, which was calculated as $M_p^{\text{FSS}}=0.94$. The model also accounted for an additional reduction in polarization mode overlap due to setup birefringence (induced by imperfect polarization control), and non-perfect temporal mode overlap due to a finite temporal resolution. Given the uncertainty in the impact of setup birefringence on $M_p$, the polarization mode overlap was assumed to be within the interval $M_p=\left[ 0.8, 0.9 \right]$. It is important to note that the additional $\ketbra{H}{H}$ term in the teleportation output state is specific to the chosen alignment procedure, representing a worst-case scenario. Here, the TPI visibility is maximized only for one of the polarized wave packets—in this case, the $\ket{H}$ polarized packet. 

The TPI visibility was calculated for individual time window sizes from the measurement data, see Fig.~\ref{fig:fig2}. Similarly, the ratio of unwanted three-fold coincidences was derived by processing the measured $g^{(2)}(0)$ for all three interacting photons in individual time windows. From the dependencies of individual $g^{(2)}(0)$ on the time window size the two-photon component was estimated as~\cite{Vyvlecka2023}

\begin{equation}
\label{eq:pm_bound_g2}
    p_\text{2} \le \frac{1 - B g^{(2)}(0) - \sqrt{1 - 2Bg^{(2)}(0)}}{g^{(2)}(0)} \ ,
\end{equation}
where $B$ is the QD single-photon source brightness at the detector and the assumption of $p_{\ge 3}=0$ is used (see supplement). All possible combinations of detector clicks leading to unwanted three-fold coincidences are accounted for. The ratio of wanted coincidences to all detected coincidences is then, for a $\SI{70}{\pico\second}$ time window, given as $k_{\SI{70}{\pico\second}}=0.85$.

\medskip

\large

\noindent\textbf{Data availability}

\normalsize

\noindent The data generated and analyzed during the current study is available from the corresponding author upon reasonable request.

\medskip

\large

\noindent\textbf{Funding}

\normalsize

\noindent This work is funded by the German Federal Ministry of Education and Research (BMBF) via the projects QR.X (Contracts No.~16KISQ013, 16KISQ001K and 16KISQ016) and Q.Link.X (Contract No.~16KIS0864). This project has received funding from the European Union’s Horizon 2020 research and innovation program under Grant Agreement no. 899814 (Qurope).

\medskip

\large

\noindent\textbf{Acknowledgments}

\normalsize

\noindent The authors gratefully acknowledge the company Single Quantum for their persistent support. We further thank Montana Instruments and Quantum Design for their support with the cryostats. We also thank Nam Tran for his contribution in creating the 3D graphics.

\medskip

\large

\noindent\textbf{Competing interests}

\normalsize

\noindent The Authors declare no Competing Financial or Non-Financial Interests.

\large

\medskip

\noindent\textbf{Author contributions}

\normalsize

\noindent T.S. and I.N. carried out the experiments with the support of S.K., R.J., and C.H.. T.B. and M.S. set up the QFC devices for the experiment and supported the experimental preparations. M.V. performed the theoretical modeling. T.S., M.V., and R.J. analyzed the data. J.H.W. designed and constructed the micro-photoluminescence setup, assisted by T.S. and S.L.P.. C.N. designed and built the BSM setup. N.S. grew the samples, which were processed and pre-characterized by W.N. and G.B.. T.S. and S.K. pre-selected QD candidates. T.S., M.V., and S.L.P. drafted the manuscript and all authors contributed to the final version. C.H., C.B., P.M., and S.L.P. coordinated the project.

\normalsize

\medskip

\large

\bibliography{Nature_Submission}

\onecolumngrid
%\vspace{\columnsep}
%\lipsum[5-6]
%\vspace{\columnsep}
%-------------------------------------------------------

\end{document}